\documentclass[useAMS,usenatbib]{mn2e}
\usepackage{graphicx}
\usepackage{color}
\usepackage{amsmath}

\title[The emitting areas in the nucleus of NGC 1313]{The emission-line regions in the nucleus of NGC 1313 probed with GMOS-IFU: a supergiant/hypergiant candidate and a kinematically cold nucleus}

\author[Menezes \& Steiner] {R.~B.~Menezes$^{1,2}$\thanks{E-mail:
robertobm@astro.iag.usp.br} and J.~E.~Steiner$^1$ \\
$^{1}$Instituto de Astronomia Geof\'isica e Ci\^encias Atmosf\'ericas, Universidade de S\~ao Paulo, Rua do Mat\~ao 1226, \\Cidade Universit\'aria, S\~ao Paulo, SP CEP 05508-090, Brazil \\
$^{2}$IP\&D, Universidade do Vale do Para\'iba, Av. Shishima Hifumi, 2911, S\~ao Jos\'e dos Campos, SP CEP 12244-000, Brasil}

\begin{document}

\date{Submitted 2016 February 27}

\pagerange{\pageref{firstpage}--\pageref{lastpage}} \pubyear{2017}

\maketitle

\label{firstpage}

\begin{abstract}

NGC 1313 is a bulgeless nearby galaxy, classified as SB(s)d. Its proximity allows high spatial resolution observations. We performed the first detailed analysis of the emission-line properties in the nuclear region of NGC 1313, using an optical data cube obtained with the Gemini Multi-object Spectrograph. We detected four main emitting areas, three of them (regions 1, 2 and 3) having spectra typical of H II regions. Region 1 is located very close to the stellar nucleus and shows broad spectral features characteristic of Wolf-Rayet stars. Our analysis revealed the presence of one or two WC4-5 stars in this region, which is compatible with results obtained by previous studies. Region 4 shows spectral features (as a strong H$\alpha$ emission line, with a broad component) typical of a massive emission-line star, such as a luminous blue variable, a B[e] supergiant or a B hypergiant. The radial velocity map of the ionized gas shows a pattern consistent with rotation. A significant drop in the values of the gas velocity dispersion was detected very close to region 1, which suggests that the young stars there were formed from this cold gas, possibly keeping low values of velocity dispersion. Therefore, although detailed measurements of the stellar kinematics were not possible (due to the weak stellar absorption spectrum of this galaxy), we predict that NGC 1313 may also show a drop in the values of the stellar velocity dispersion in its nuclear region. 

\end{abstract}


\section{Introduction}

Late-type galaxies usually have low mass, a large amount of gas and dust and show significant star formation. Their spheroidal component, the bulge, is much less prominent than in the case of early-type galaxies. An extreme class of these objects is composed of the bulgeless galaxies (with morphological types from Scd to Sm), which have essentially no bulge \citep{bok02}. Many of the processes related to the formation and evolution of these objects are not completely understood. NGC 1313 is a member of this extreme class of late-type galaxies, classified as SB(s)d by \citet{vau63}, who mentioned that this object seems to be a transition between the morphological types SBc and SBm. It has almost symmetric spiral arms, although the northern arm is a little more developed than the southern one. Being located at a distance of only 4.5 Mpc, this is an ideal target for morphological and spectroscopic studies of bulgeless galaxies with large telescopes.

\begin{figure*}
\begin{center}
  \includegraphics[scale=0.40]{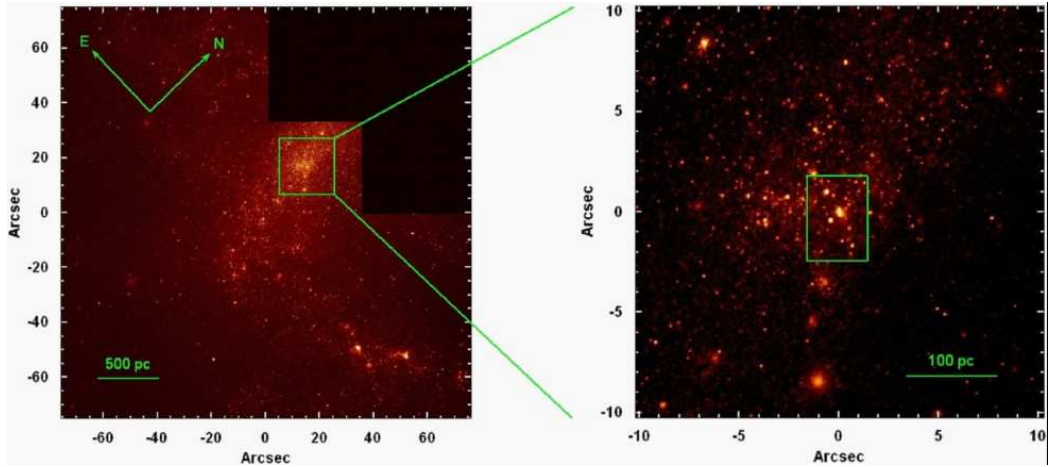}
  \caption{Left: HST image of NGC 1313, obtained with WFPC2 in the F606W filter. Right: magnification of the highlighted section of the image shown at left, with our GMOS/IFU pointing indicated.\label{fig1}}
\end{center}
\end{figure*}

Using a ground based H$\alpha$ image, \citet{mar83} elaborated a catalogue of 375 H II regions in NGC 1313, some of them very close to the nucleus. The authors concluded that the observed H II regions have an average diameter of 18.4 pc, the brightest ones being located along the spiral arms. \citet{pag80} analysed spectrophotometric data of six H II regions in NGC 1313 and verified that the abundance distribution is very uniform and can be given by $12 + log(O/H) = 8.26 \pm 0.07$. Such abundance is similar to that of the Large Magellanic Cloud ($8.37 \pm 0.22$; Russell \& Dopita 1990). In a latter study, \citet{wal97} presented multi-fibre spectrophotometry of 33 H II regions in this galaxy and obtained a similar abundance distribution for the disc, given by $12 + log(O/H) = 8.4 \pm 0.1$, the bar possibly having a higher abundance than the disc by a factor of 0.2 dex.  

\citet{had07} performed a survey of Wolf-Rayet stars (WRs) in NGC 1313, using optical data obtained with the \textit{Very Large Telescope}, and detected 70 objects with spectral features characteristic of WRs, one of them nearly coincident with the nucleus of the galaxy. The observed nebular properties resulted in an abundance distribution of $12 + log(O/H) = 8.23 \pm 0.06$. In addition, using continuum-subtracted H$\alpha$ images of this galaxy, the authors estimated a global star formation rate of 0.6 $M_{\sun}$ yr$^{-1}$.

The first kinematic study of NGC 1313 was made by \citet{car74}, who analysed the kinematics of the H$\alpha$ emission line and verified that the rotation centre does not coincide with the nucleus. \citet{mar82} mapped the velocity field of the ionized gas in this galaxy, using interferometric data of H$\alpha$, and concluded that the rotation centre is located outside the bar, at a distance of 1.5 kpc from the nucleus. However, a detailed analysis of the H I flux and kinematics by \citet{pet94} revealed that most of the velocity field can be interpreted as a simple circular rotation around the nucleus, with no displacement of the rotation centre. \citet{pet94} also detected a tidal interaction with a possible satellite galaxy, which was recently disrupted.

\begin{figure*}
\begin{center}
  \includegraphics[scale=0.59]{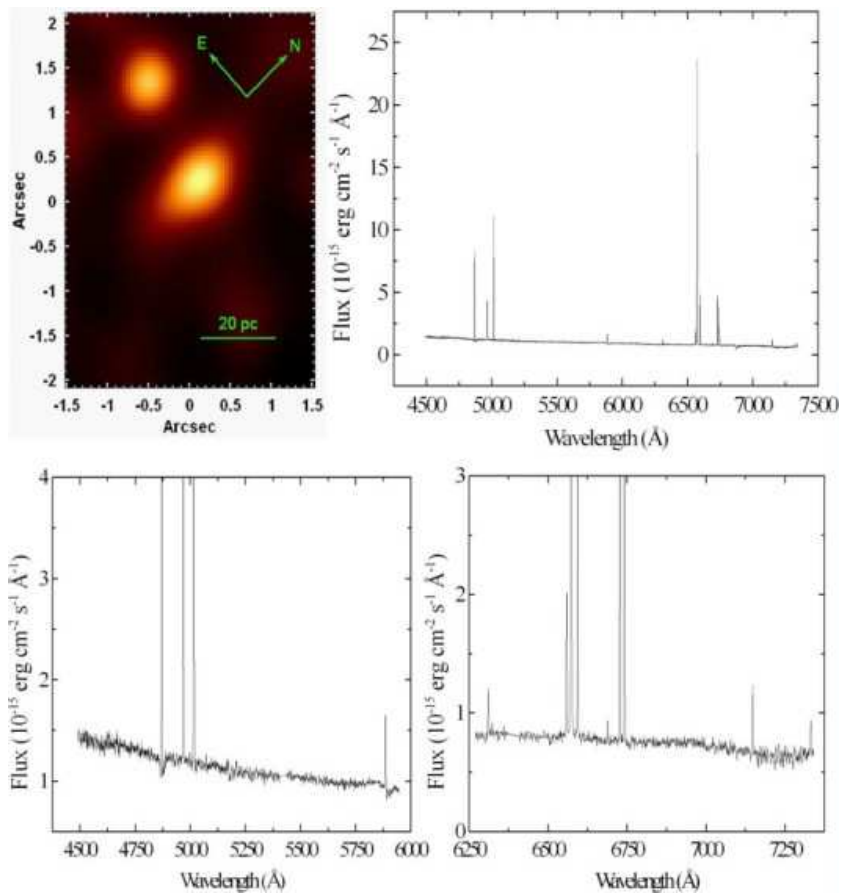}
  \caption{Average spectrum and image of the collapsed data cube of NGC 1313, after the data treatment. Magnifications of different wavelength ranges of the average spectrum are shown at the bottom.\label{fig2}}
\end{center}
\end{figure*}

\citet{sil11} analysed the star formation history of this galaxy, using high-resolution images, obtained with the \textit{Hubble Space Telescope} (\textit{HST}), and determined star formation rate densities of $(14.9 - 30.9) \times 10^{-3}$ M$_{\sun}$ yr$^{-1}$ kpc$^{-2}$ for the whole galaxy. In addition, the authors detected evidence of a recent starburst south-west from the nucleus, which is compatible with the hypothesis of a tidal interaction with a disrupted satellite galaxy.

Besides the presence of H II regions \citep{mar83} and WRs \citep{had07}, there is not much information in the literature about the nuclear region of NGC 1313. Our research group is currently conducting the \textit{Deep IFS View of Nuclei of Galaxies} ($DIVING^{3D}$) survey, whose goal is to observe the nuclear regions of all galaxies in the Southern hemisphere brighter than $B = 12.0$. The principal investigator of this project is JES. The observations are being taken, in the optical, with the integral field unit (IFU) of the Gemini Multi-object Spectrograph (GMOS), mounted at the Gemini-south telescope. NGC 1313 is one of the brightest bulgeless galaxies in our sample and, therefore, deserves special attention. In this paper, we present the first detailed study of the emission-line properties in the nuclear region of NGC 1313, based on an optical data cube obtained with the GMOS-south/IFU, as part of the $DIVING^{3D}$ survey. In Section 2 we describe the observations, the data reduction and the data treatment. The results of our analysis are presented in Section 3. In Section 4 we discuss and compare our results with previous studies. Finally, we summarize our work and present our conclusions in Section 5.

\section{Observations and data reduction}

The observations of the nuclear region of NGC 1313, with coordinates RA = 03\textsuperscript{h} 18\textsuperscript{m} 16.0\textsuperscript{s} and Dec. = -66$\degr$ 29$\arcmin$ 54$\arcsec\!\!.0$, were taken on 2012 December 4. Three 590 s exposures, with spatial dithering of 0.2 arcsec, were obtained. As we used the IFU in the one-slit mode, the science field of view (FOV) of the        observations (sampled by 500 fibres) has 5 arcsec $\times$ 3.5 arcsec and the sky FOV (located at a distance of 1 arcmin from the science FOV and sampled by 250 fibres) has 5 arcsec $\times$ 1.75 arcsec. The observations were taken with the B600+G5323 grating, which resulted in a spectral coverage of $4500 - 7350 \AA$ and a spectral resolution of $R \sim 3085$. Based on acquisition images, we estimated a seeing of $\sim 0.64$ arcsec. Fig.~\ref{fig1} shows an \textit{HST} image of NGC 1313, obtained with the Wide-field Planetary Camera 2 (WFPC2) in the F606W filter, with our IFU pointing indicated.  

The required calibration images (bias, GCAL flat, twilight flat and CuAr lamp) were obtained during the observations. In order to perform flux calibration of the data, the standard star LTT 1020 was also observed, on 2012 November 28. The data reduction was performed using the Gemini \textsc{IRAF} package. The first part of the data reduction consisted of the trim and bias subtraction of the data. After that, the cosmic rays were removed with the L.A.Cosmic routine \citep{van01}. Then, the spectra were extracted, corrected for gain variations along the spectral axis (using the response curves provided by the GCAL flat images), corrected for gain variations along the FOV and for illumination patterns of the instrument (with response maps obtained with the twilight flat) and wavelength calibrated. The average spectrum obtained with the sky FOV was subtracted from science data. It is important to mention, however, that, being located at 1 arcmin from the nucleus of NGC 1313, the sky FOV could be contaminated by the emission from the galaxy's disc. However, a careful inspection of the sky spectrum revealed no emission lines other than the sky lines. In addition, we performed a cross-correlation between the sky spectrum and an average spectrum of the science data and found no indication of a significant contamination of the sky spectrum by the emission from the galaxy's disc. Finally, the telluric absorptions were removed, the data were flux calibrated (with also a correction for the atmospheric extinction) and the data cubes were constructed, with spatial pixels (spaxels) of 0.05 arcsec.

\begin{figure*}
\begin{center}
  \includegraphics[scale=0.35]{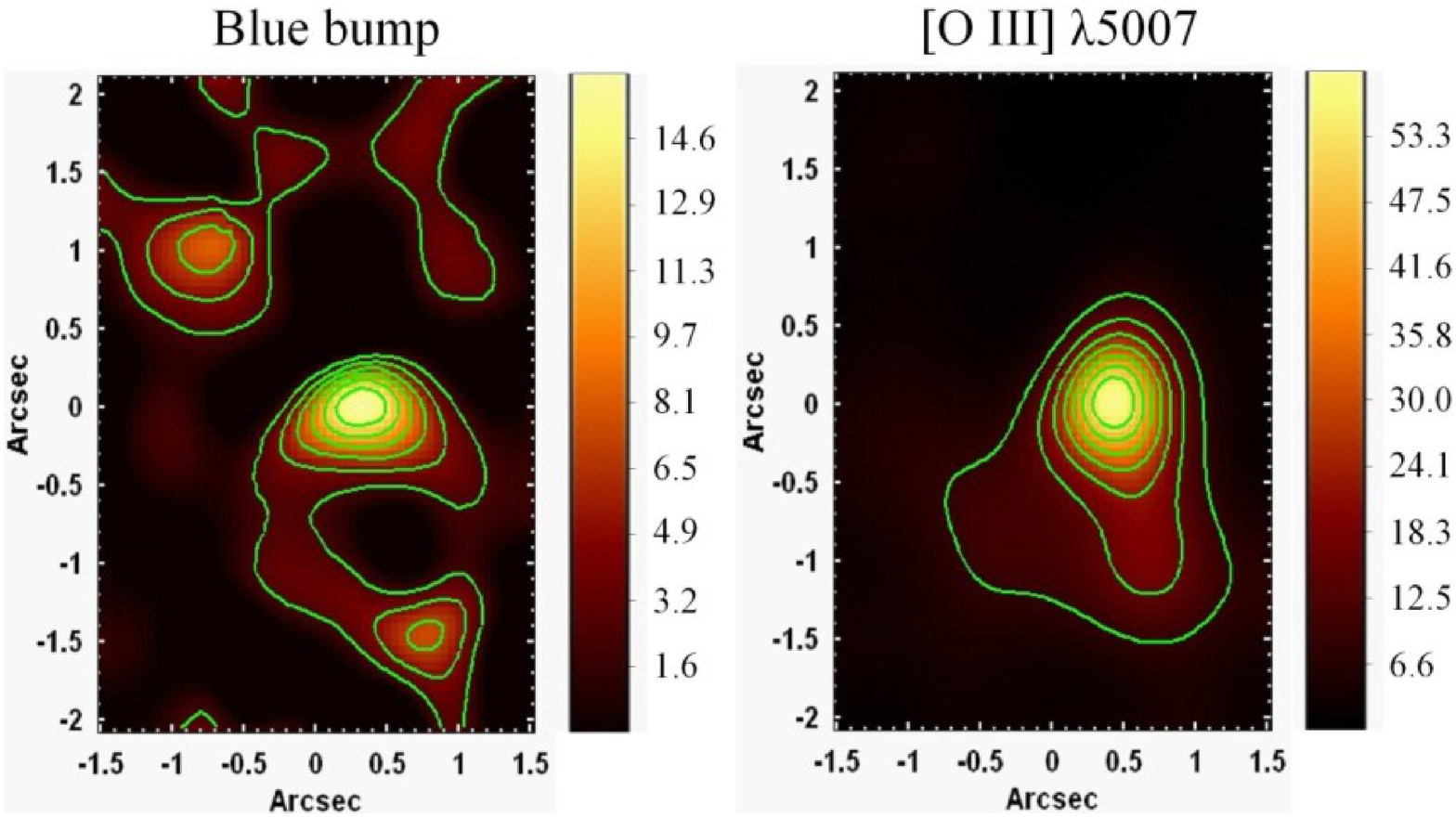}
  \includegraphics[scale=0.35]{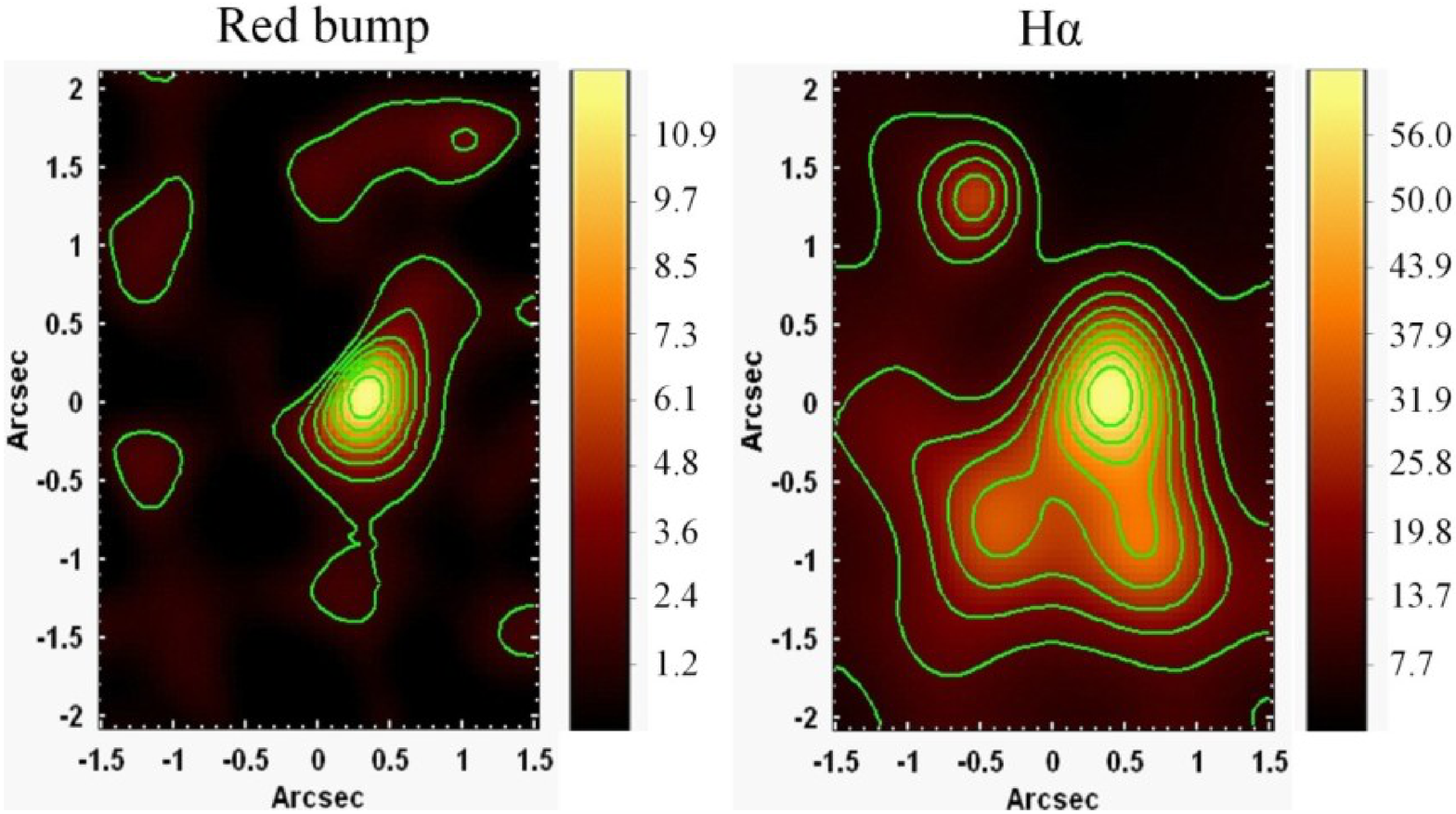} \\
  \includegraphics[scale=0.35]{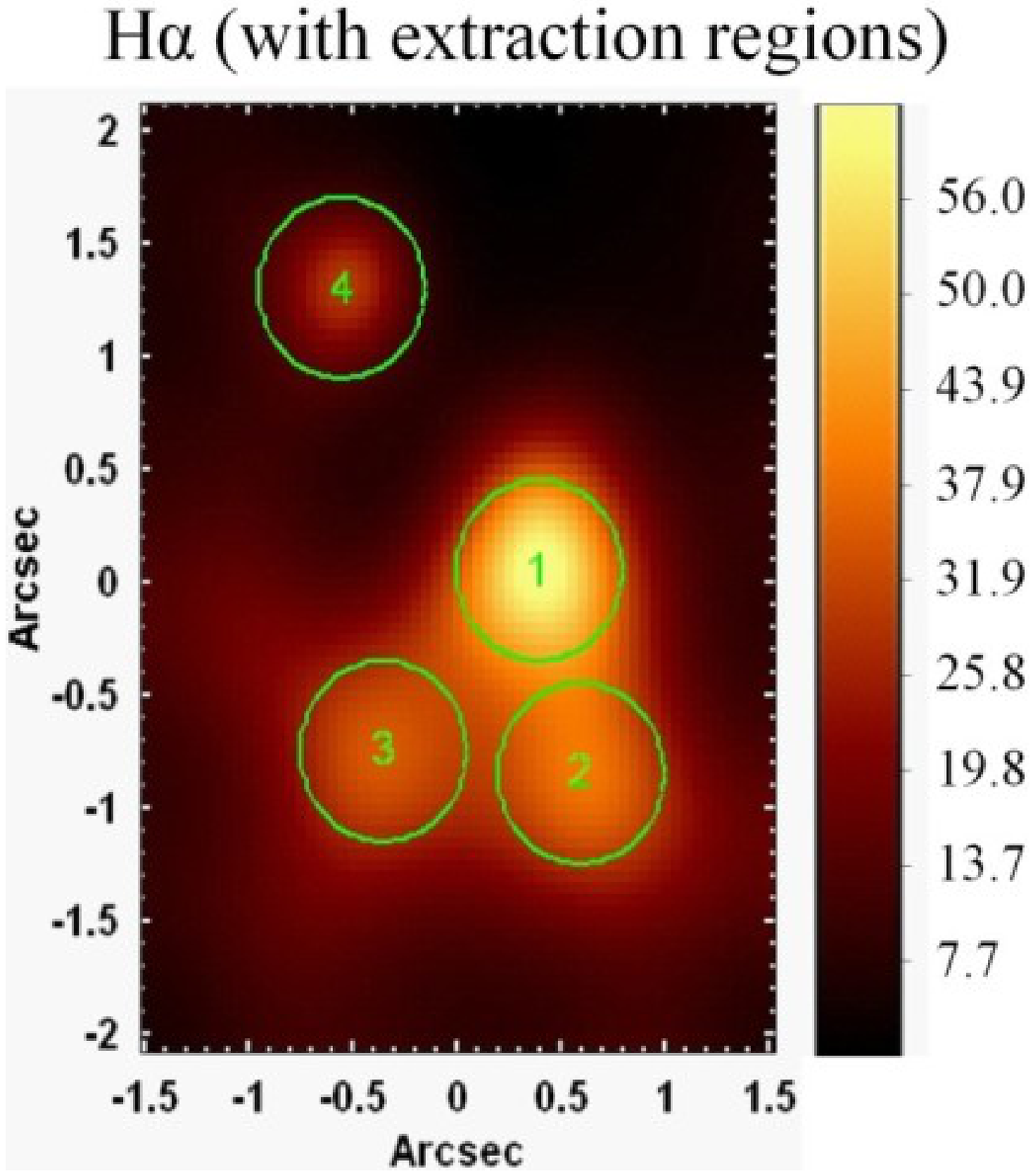}
  \caption{Integrated flux maps of the main emission features in the data cube of NGC 1313, after the starlight subtraction. The isocontours are shown in green. The four circular regions from which spectra were extracted are superposed to the H$\alpha$ image at the bottom. All the flux values are in $10^{-15}$ erg cm$^{-2}$ s$^{-1}$.\label{fig3}}
\end{center}
\end{figure*}

We treated the reduced data cubes with the procedure described in Menezes, Steiner \& Ricci (2014) and \citet{men15}. First, the differential atmospheric refraction effect was removed from the data cubes. Then, a median of the data cubes was calculated. After that, we removed high spatial-frequency noise from the images of the data cubes, using the Butterworth spatial filtering \citep{gon02}, and also removed an instrumental fingerprint (which had a spectral signature and appeared in the form of vertical stripes across the images), using the Principal Component Analysis (PCA) Tomography technique \citep{ste09}. Finally, in order to improve the spatial resolution of the observation, we applied a Richardson-Lucy deconvolution \citep{ric72, luc74} to all the images of the data cube. 

The Richardson-Lucy deconvolution requires an estimate of the original point-spread function (PSF) of the data cube at a certain wavelength. As explained in Section 3, this data cube has a point-like source, east from the nucleus, which is a supergiant/hypergiant star candidate. The spectrum of this source shows a strong H$\alpha$ emission line, with a broad component. An image of the blue wing of this broad component (after the subtraction of an image of the adjacent stellar continuum) provided a reliable estimate of the PSF at $6568\AA$, which can be well described by a Gaussian with a full width at half-maximum (FWHM) of $\sim 0.63$ arcsec. This value is consistent with the estimate obtained from the acquisition images and from a comparison between the image of the data cube, collapsed along the spectral axis, with a convolved \textit{V} band \textit{HST} image of this galaxy, obtained with the Advanced Camera for Surveys (ACS). It is important to emphasize that the wavelength of this reference PSF does not take into account any redshift correction. 

In order to apply the Richardson-Lucy deconvolution, it is also necessary to determine the way the FWHM of the PSF varies with the wavelength. An equation describing the variation of the PSF with the wavelength was obtained from the data cube of the standard star LTT 1020, which was used in the data reduction to perform the flux calibration. Using the image of the blue wing of the broad H$\alpha$ component, we verified that the FWHM of the PSF of the treated data cube was $\sim 0.52$ arcsec, which is, again, consistent with an estimate obtained from a comparison betwen the collapsed deconvolved data cube and a convolved \textit{V} band \textit{HST}/ACS image of this galaxy.

Fig.~\ref{fig2} shows the image of the treated data cube of NGC 1313, collapsed along the spectral axis, and its average spectrum. Two bright areas can be seen in the image, the brightest one coinciding with the stellar nucleus of the galaxy. The average spectrum reveals prominent narrow emission lines, with no apparent broad component.

\section{Data analysis and results}

Since this paper is focused on the emission-line spectrum of the data cube of NGC 1313, the first step of our analysis consisted of the starlight subtraction of the data cube. This procedure was applied using synthetic stellar spectra provided by the \textsc{Starlight} software \citep{cid05}, which performs a spectral synthesis using template stellar spectra from a pre-established base. For this work, we chose a base of stellar population spectra based on Medium-resolution Isaac Newton Telescope Library of Empirical Spectra (MILES; S\'anchez-Bl\'azquez et al. 2006). This base has a spectral resolution of FWHM = $2.3 \AA$, which is close to our spectral resolution (FWHM = $1.8 \AA$). The spectra of the data cube were corrected for the Galactic extinction, using $A_V = 0.299$ mag (NASA Extragalactic Database - NED) and the reddening law of \citet{car89}, and also shifted to the rest frame, using $z = 0.001568$ (NED), before the spectral synthesis was applied. 

\begin{figure*}
\begin{center}
  \includegraphics[scale=0.25]{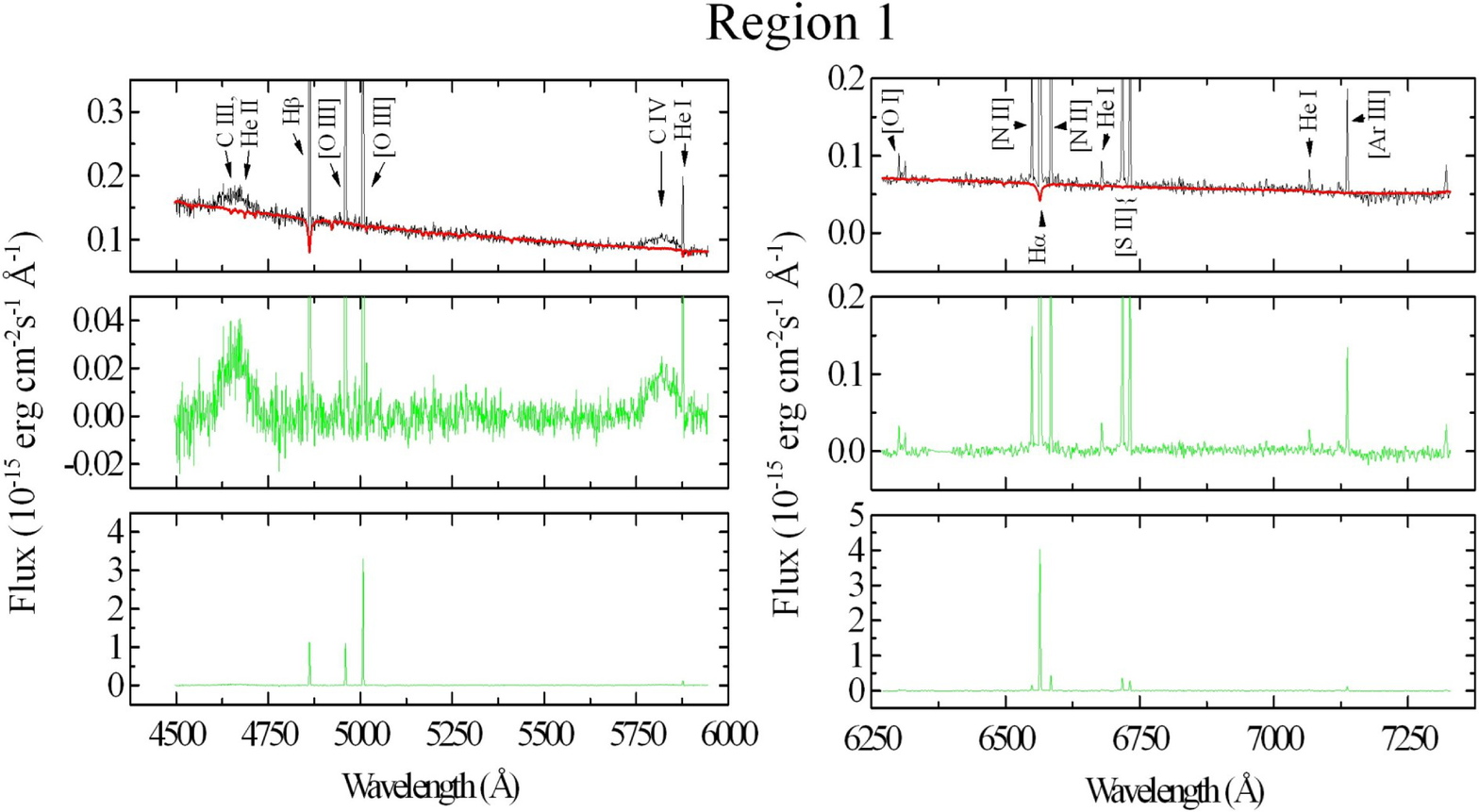}
  \includegraphics[scale=0.25]{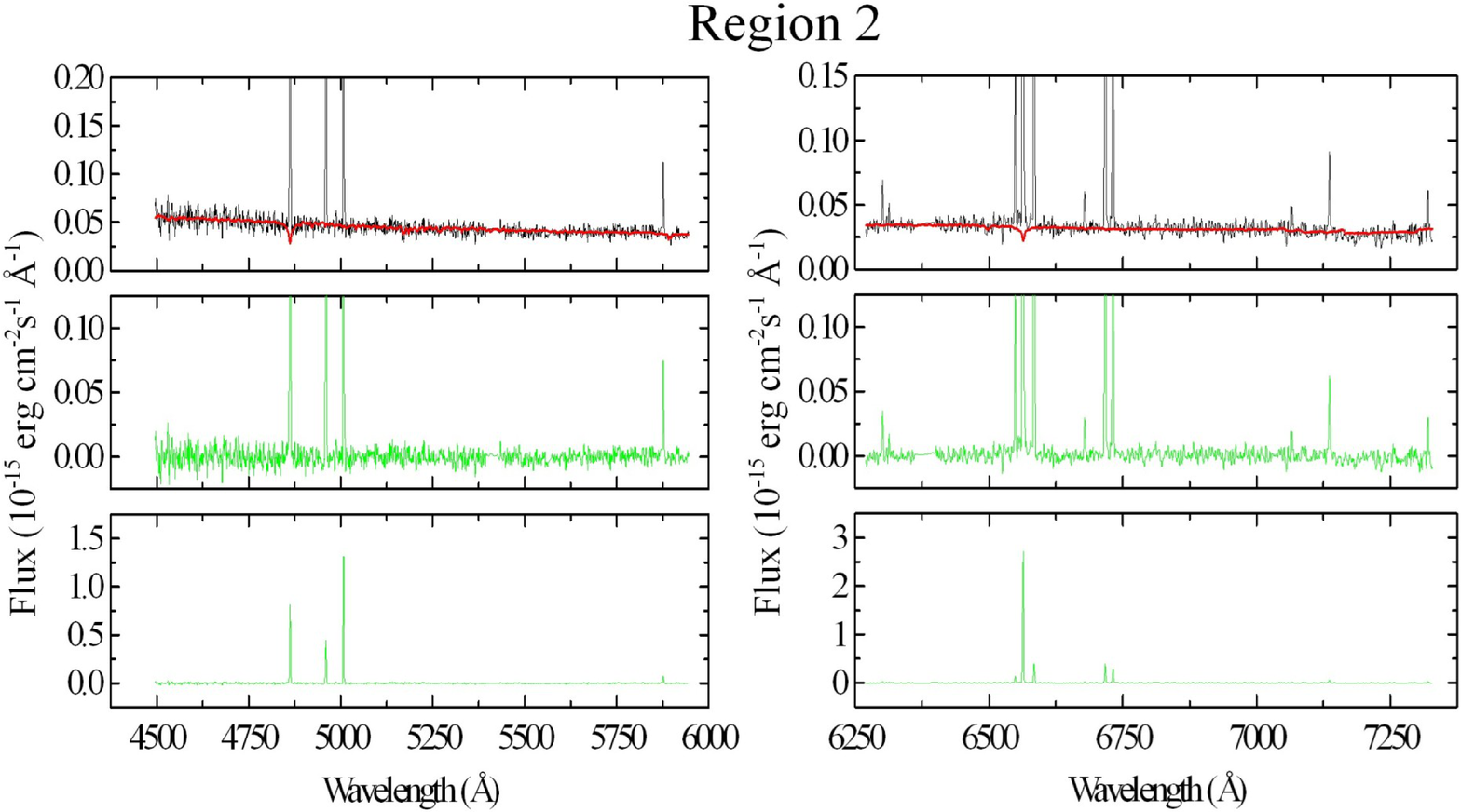}
  \caption{Different magnifications and wavelength ranges of the spectra extracted from regions 1 and 2 of the data cube of NGC 1313. The circular regions from which the spectra were extracted have a radius of 0.4 arcsec. The synthetic stellar spectra provided by the spectral synthesis are shown in red and the emission-line spectra obtained by subtracting the synthetic spectra from the observed ones are shown in green.\label{fig4}}
\end{center}
\end{figure*}

\begin{figure*}
\begin{center}
  \includegraphics[scale=0.25]{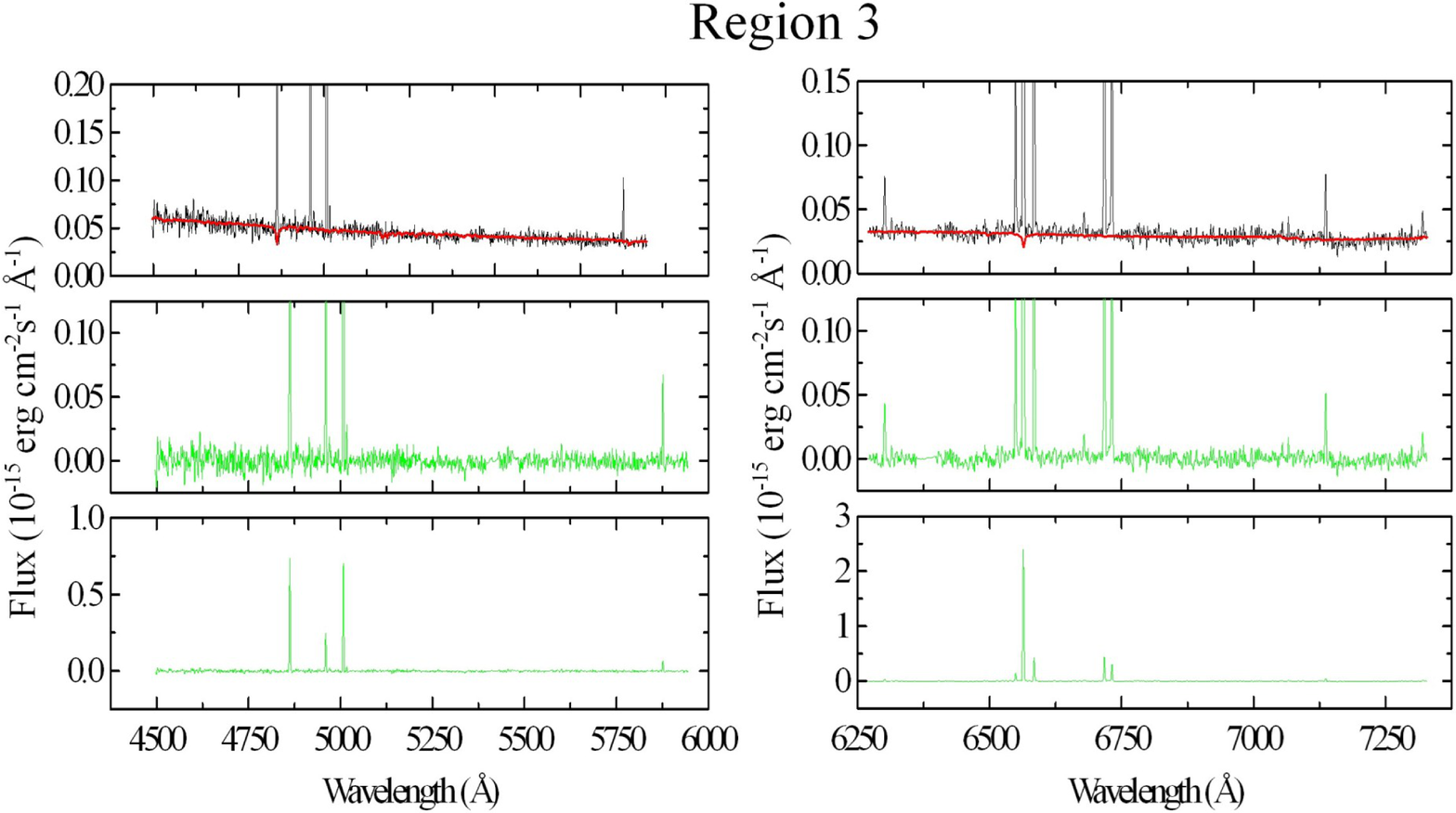}
  \includegraphics[scale=0.35]{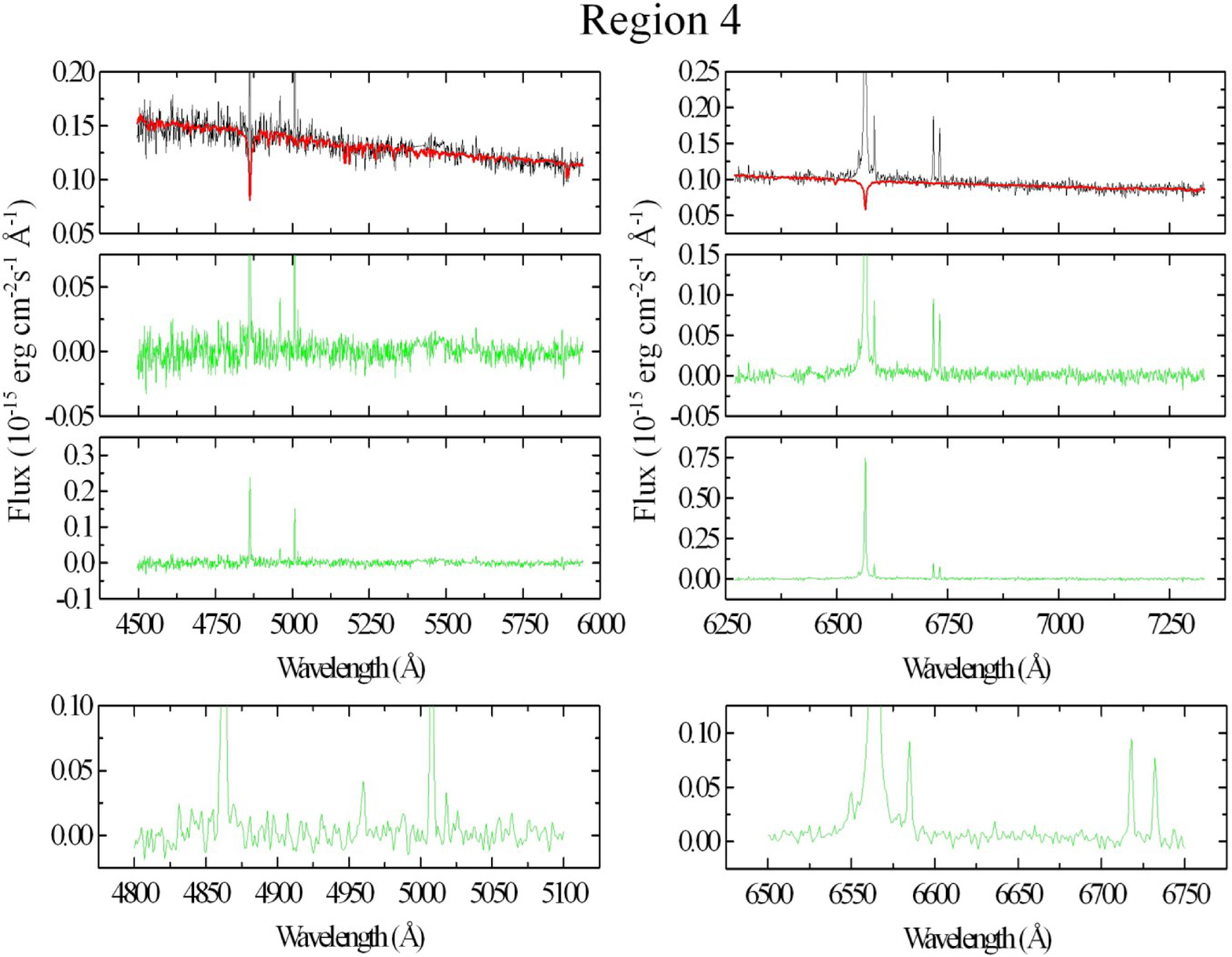}
  \caption{The same as Fig.~\ref{fig4} but for the spectra extracted from regions 3 and 4. Two additional magnifications of the emission-line spectrum of region 4 are shown at the bottom.\label{fig5}}
\end{center}
\end{figure*}

\begin{figure*}
\begin{center}
  \includegraphics[scale=0.50]{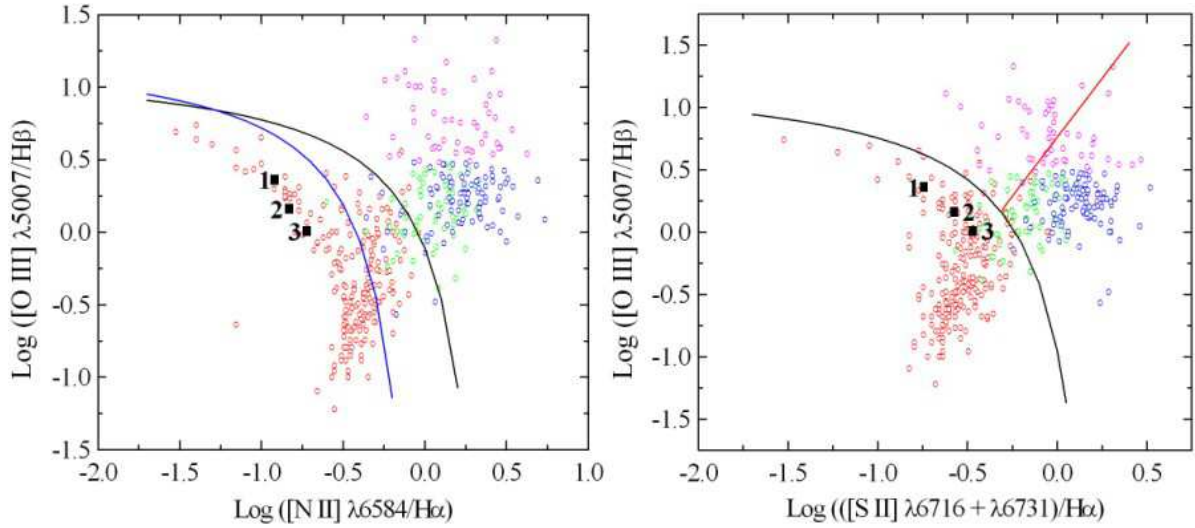}
  \caption{Diagnostic diagrams with the points representing regions 1, 2 and 3 of the data cube of NGC 1313. The other points correspond to objects analysed by Ho et al. (1997). The open circles in red, green, blue and magenta represent H II regions, transition objects, low ionization nuclear emission-line regions (LINERs) and Seyfert galaxies, respectively. The blue curve represents an empirical division between H II regions and AGNs \citep{kau03}, the black curve corresponds to the maximum limit for the ionization by a starburst \citep{kew01} and the red curve represents a division between LINERs and Seyfert galaxies \citep{kew06}.\label{fig6}}
\end{center}
\end{figure*}

\begin{table*}
\begin{center}
\caption{Emission-line ratios, luminosities of the H$\alpha$ emission line and projected distances from the stellar nucleus ($d$) of regions 1, 2 and 3 of the data cube of NGC 1313.\label{tbl1}}
\begin{tabular}{cccc}
\hline
Ratio & 1 & 2 & 3 \\
\hline
$[$N II$]$ $\lambda 6584$/H$\alpha$ & $0.121 \pm 0.009$ & $0.148 \pm 0.010$ & $0.190 \pm 0.013$ \\
($[$S II$]$ $\lambda 6716 + \lambda 6731$)/H$\alpha$ & $0.181 \pm 0.012$ & $0.267 \pm 0.016$ & $0.339 \pm 0.021$ \\
$[$O III$]$ $\lambda 5007$/H$\beta$ & $2.29 \pm 0.16$ & $1.44 \pm 0.10$ & $1.02 \pm 0.07$ \\
$L_{H\alpha}$ ($10^{37}$ erg s$^{-1}$) & $3.05 \pm 0.15$ & $1.60 \pm 0.08$ & $1.417 \pm 0.07$ \\
$d$ (arcsec) & $0.36 \pm 0.06$ & $1.21 \pm 0.06$ & $1.10 \pm 0.06$ \\
\hline
\end{tabular}
\end{center}
\end{table*}

By subtracting the synthetic stellar spectra provided by the spectral synthesis from the observed ones, we obtained a data cube with only emission lines. In this residual data cube, besides the narrow emission lines visible in the average spectrum of the original data cube (Fig.~\ref{fig2}), we also detected two broad features: a ``blue bump'' representing a blend of emission lines around $4650 \AA$ (including lines such as N III $\lambda \lambda \lambda 4628, 4634, 4640$, C III $\lambda 4650$, C IV $\lambda 4658$ and He II $4686$) and a ``red bump'' representing the C IV $\lambda \lambda 5801, 5812$ lines. Fig.~\ref{fig3} shows integrated flux maps of the blue and red bumps and of the [O III] $\lambda 5007$ and H$\alpha$ emission lines, with isocontours. The blue and red bumps are emitted in a compact area, whose position is compatible with the position of the brightest region in the H$\alpha$ and [O III] $\lambda 5007$ images. The H$\alpha$ emission comes mainly from four areas, the brightest of them located west and at a projected distance of $0.36 \pm 0.06$ arcsec (which corresponds to $7.9 \pm 1.3$ pc) from the stellar nucleus. The spatial morphologies of the [O III] $\lambda 5007$ and H$\alpha$ emission lines are relatively similar to each other. The main difference between the images of these lines is an isolated emitting region, located east and at a projected distance of $1.24 \pm 0.06$ arcsec ($26.9 \pm 1.3$ pc) from the stellar nucleus, which is clearly visible in H$\alpha$ but does not appear in [O III] $\lambda 5007$. 

We extracted the spectra from four circular regions of the data cube of NGC 1313, before the starlight subtraction, centred on each one of the H$\alpha$ emitting areas. The circular regions have a radius of 0.4 arcsec and are illustrated in an H$\alpha$ image in Fig.~\ref{fig3}. Figs. 4 and 5 show the extracted spectra, together with the fits provided by the spectral synthesis and the fit residuals. The blue and red bumps discussed above can be easily seen in the spectrum of region 1. These broad features are typically seen in the spectra of WRs, which indicates the presence of these massive evolved stars in region 1. The values of the FWHM of the blue and red bumps (corrected for the spectral resolution) are FWHM$_{blue~bump} = 4900 \pm 150$ km s$^{-1}$ and FWHM$_{red~bump} = 4200 \pm 200$ km s$^{-1}$, respectively. The spectra of regions 2 and 3 are similar to each other and reveal only narrow emission lines. On the other hand, the spectrum of region 4 is considerably different from the others, as it shows an apparent broad component of the H$\alpha$ line. Based on its characteristics, we believe that the spectrum of region 4 is due to the presence of a massive emission-line star (like a luminous blue variable - LBV, a B[e] supergiant or a B hypergiant) in this area (see the discussion in Section 4). 

We corrected the spectra of regions 1, 2, 3 and 4 for the interstellar extincton (in the observed galaxy), using the $A_V$ values provided by the spectral synthesis of these spectra and also the reddening law of Cardelli et al. (1989). After this correction, we verified that the Balmer decrements (H$\alpha$/H$\beta$) of the extracted spectra were all compatible, at 1$\sigma$ level, with 2.86, which is the expected value for Case B recombination, considering an electron density of $10^2$ cm$^{-3}$ and a temperature of $10^4$ K \citep{ost06}. We calculated the luminosities of the blue and red bumps for the spectrum of region 1 (assuming a distance of 4.5 Mpc for the galaxy) and obtained $L_{blue~bump} = (7.7 \pm 0.4) \times 10^{36}$ erg s$^{-1}$ and $L_{red~bump} = (5.17 \pm 0.26) \times 10^{36}$ erg s$^{-1}$. We also calculated the emission-line ratios [N II] $\lambda 6584$/H$\alpha$, ([S II] $\lambda 6716 + \lambda 6731$)/H$\alpha$ and [O III] $\lambda 5007$/H$\beta$ for the spectra of regions 1, 2 and 3. Table~\ref{tbl1} shows the emission-line ratios, together with the luminosities of the H$\alpha$ line and the projected distances between each region and the stellar nucleus. 

We constructed the diagnostic diagrams [O III] $\lambda 5007$/H$\beta$ $\times$ [N II] $\lambda 6584$/H$\alpha$ and [O III] $\lambda 5007$/H$\beta$ $\times$ ([S II] $\lambda 6716 + \lambda 6731$)/H$\alpha$ and included the points representing regions 1, 2 and 3. In these same diagrams, we also included the points corresponding to the objects analysed by \citet{ho97}. The results are shown in Fig.~\ref{fig6}. It is easy to see that the analysed areas are H II regions, with a decreasing ionization degree from region 1 to 3. The values of the emission-line ratio [S II] $\lambda 6716$/[S II] $\lambda 6731$, along the entire FOV, were consistent with 1.44, which is the low-density limit of this ratio, assuming a temperature of $10^4$ K. This indicates that the electron densities along the FOV must satisfy $n_e < \sim 10$ cm$^{-3}$.

\begin{figure*}
\begin{center}
  \includegraphics[scale=0.50]{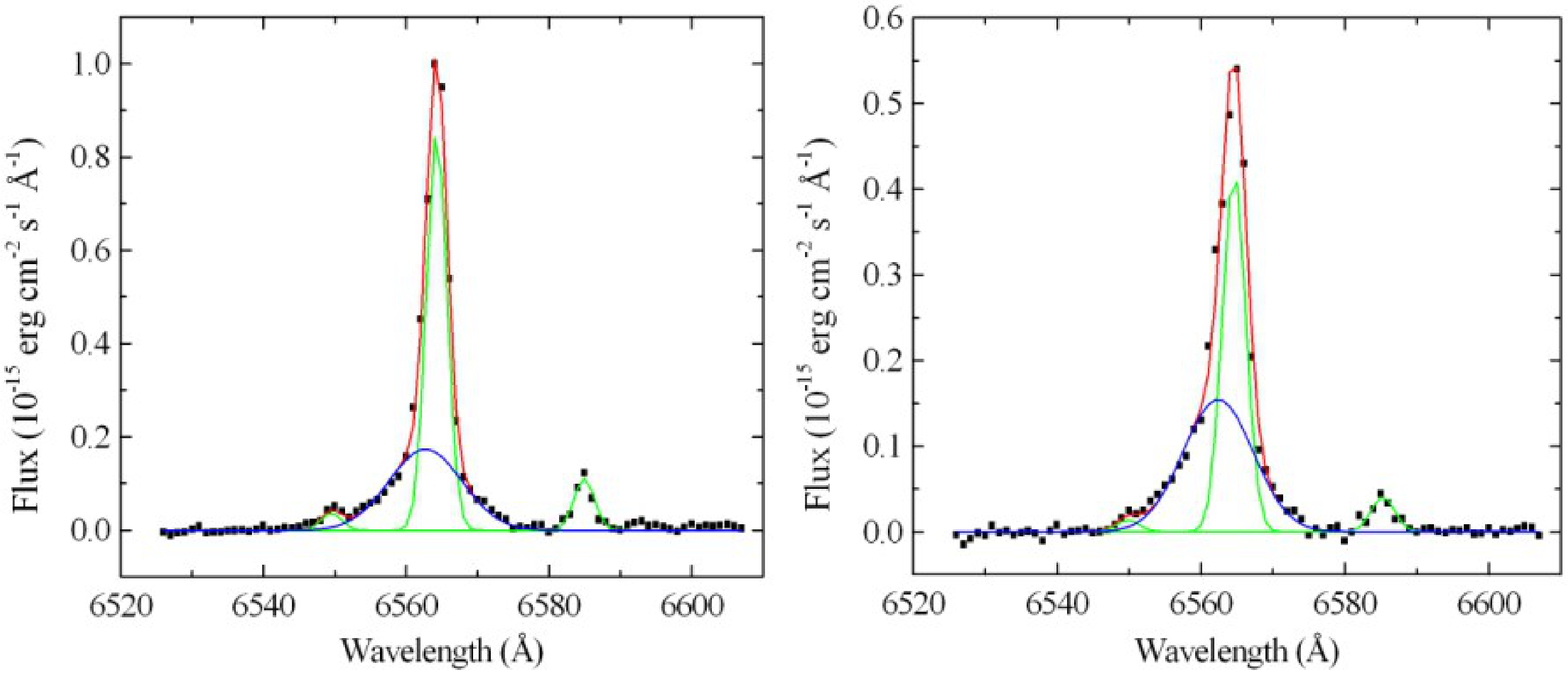}
  \caption{Gaussian fit of the [N II] + H$\alpha$ emission lines of the spectrum of region 4 before (left) and after (right) the background subtraction.  The black points represent the observed values. The narrow Gaussians used in the fit are shown in green and the broad Gaussian is shown in blue. The final fit is shown in red.\label{fig7}}
\end{center}
\end{figure*}

\begin{figure*}
\begin{center}
  \includegraphics[scale=0.65]{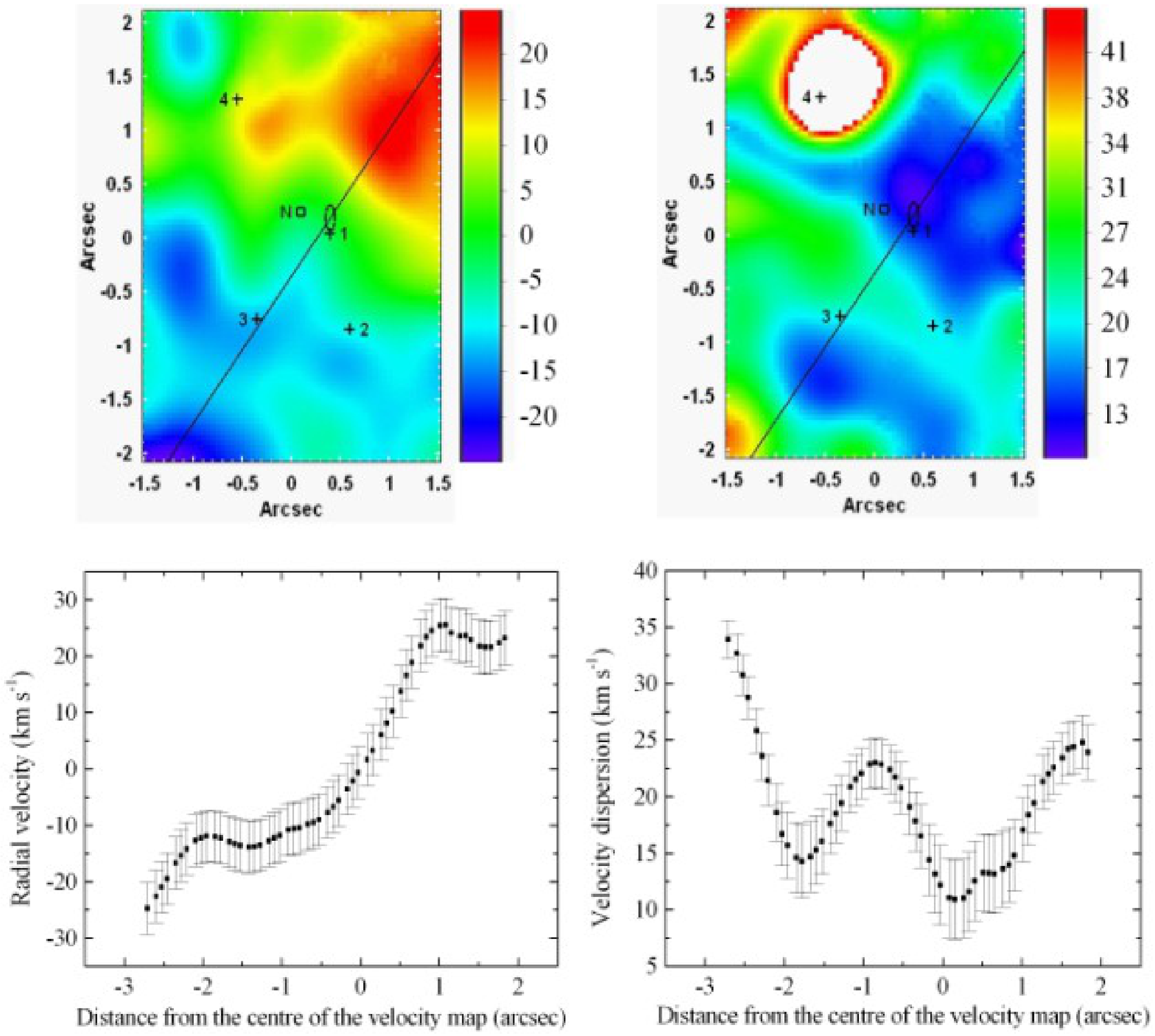}
  \caption{$V_{gas}$ map (top left) and $\sigma_{gas}$ map (top right) obtained with simple Gaussian fits of the H$\alpha$ emission line in the data cube of NGC 1313. In both maps, the positions of the four emission-line regions are marked with crosses, the position of the stellar nucleus is marked with a square and the $V_{gas}$ centre is marked with an ellipse. The half-sizes of the crosses and of the square and the semi-axis of the ellipse indicate the uncertainties ($1\sigma$) of the corresponding positions. The curves extracted, along the line of nodes, from the $V_{gas}$ map and from the $\sigma_{gas}$ map are shown at bottom left and bottom right, respectively.\label{fig8}}
\end{center}
\end{figure*}

The analysis of the spectrum of region 4 requires more caution. In order to disentangle the broad and narrow components of the H$\alpha$ line in this area, we fitted the [N II] $\lambda \lambda 6548, 6583$ and H$\alpha$ lines with a set of three narrow Gaussians, with the same width and radial velocity, and one broad Gaussian. We assumed that the intensity of the [N II] $\lambda 6548$ line is equal to 0.328 of the intensity of the [N II] $\lambda 6583$ line \citep{ost06}. The [N II] and H$\alpha$ lines were well reproduced by the fit, shown in Fig.~\ref{fig7}. The FWHM of the broad and narrow Gaussians in the fit, corrected for the instrumental resolution, are FWHM$_{broad} = 540 \pm 30$ km s$^{-1}$ and FWHM$_{narrow} = 120 \pm 6$ km s$^{-1}$, respectively. The width of the [S II] $\lambda \lambda 6716, 6731$ emission lines is compatible with the width of the narrow Gaussians used to fit the [N II] and H$\alpha$ lines. The luminosity of the broad Gaussian in the fit is $L(broad~H\alpha - 1^{st} fit) = (5.4 \pm 0.7) \times 10^{36}$ erg s$^{-1}$, while its equivalent width (calculated using the continuum of the spectrum before the starlight subtraction) is $EW(broad~H\alpha - 1^{st} fit) = -16 \pm 4 \AA$. In addition, the luminosity of the narrow component of H$\alpha$ is $L(narrow~H\alpha - 1^{st} fit) = (7.4 \pm 0.4) \times 10^{36}$ erg s$^{-1}$. 

The broad component of H$\alpha$ is originated by the probable massive emission-line star in this area. The narrow component of H$\alpha$ may be nebular; however, we cannot exclude the possibility that part of this narrow component is also emitted by the emission-line star. In order to evaluate the contribution of a possible background emission to the narrow components of the emission lines in region 4, we extracted the spectrum from an annular region, located around region 4, of the data cube of NGC 1313, after the starlight subtraction. This ``background'' was then subtracted from the spectrum of region 4, which resulted in a residual spectrum that still shows the [O III] $\lambda 5007$, [N II] $\lambda 6584$ and [S II] $\lambda \lambda 6716, 6731$ emission lines. After that, we fitted again the [N II] $\lambda \lambda 6548, 6583$ and H$\alpha$ lines with a sum of three narrow Gaussians and one broad Gaussian, considering the same assumptions mentioned before. The fit is shown in Fig.~\ref{fig7}. The widths of the lines and the integrated flux of the broad component of H$\alpha$ did not change significantly in this second fit and the new luminosity of the narrow component of H$\alpha$ ($L(narrow~H\alpha - 2^{nd} fit) = (4.2 \pm 0.4) \times 10^{36}$ erg s$^{-1}$) represents an upper limit for the narrow H$\alpha$ emitted by the star (as part of this narrow component can still be a nebular contribution emitted around the star, not removed by the background subtraction). We also applied the background subtraction to the stellar spectrum of region 4 (using spectra from the data cube before the starlight subtraction). Then, using the continuum of the resulting spectrum, we calculated the equivalent width of the broad component of H$\alpha$ provided by the second Gaussian fit and obtained a value of $EW(broad~H\alpha - 2^{nd} fit) = -20 \pm 7 \AA$. 

\begin{figure*}
\begin{center}
  \includegraphics[scale=0.50]{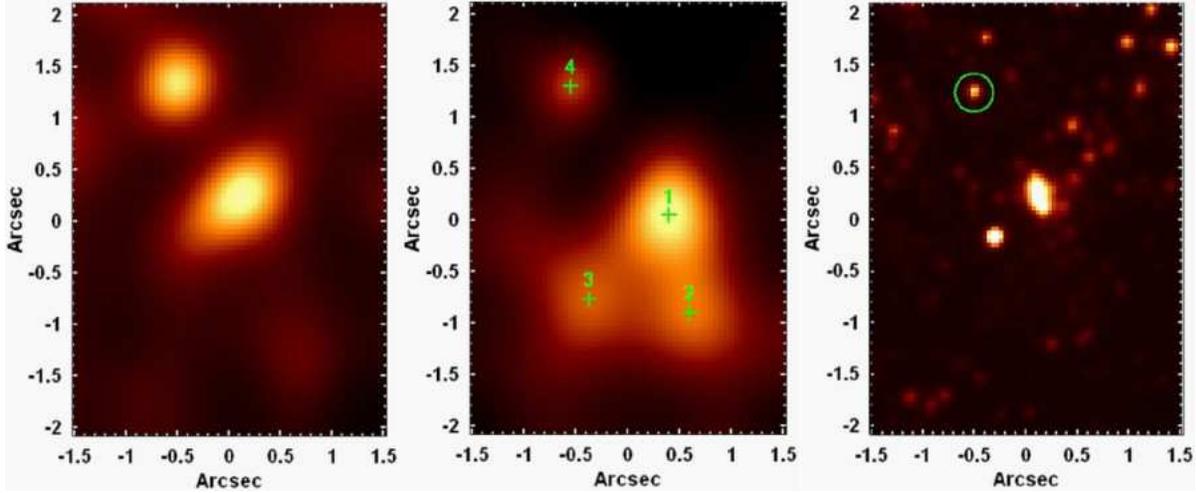}
  \caption{Left: image of the data cube of NGC 1313, before the starlight subtraction, collapsed along the spectral axis (the same shown in Fig.~\ref{fig2}). Middle: H$\alpha$ image of the data cube of NGC 1313, after the starlight subtraction (the same shown in Fig.~\ref{fig3}). The positions of the four emitting areas are marked with crosses. Right: \textit{HST} image of NGC 1313, in the F336W filter (\textit{U} band), obtained with WFC3. The LBV/B[e] supergiant/B hypergiant candidate is marked with a green circle.\label{fig9}}
\end{center}
\end{figure*}

It is important to mention that the procedure of background subtraction used here has some uncertainties. In particular, the background spectrum may be contaminated by the emission from stars not related to region 4 (see Fig.~\ref{fig9}). This certainly could affect the values of $L(narrow~H\alpha - 2^{nd} fit)$ and $EW(broad~H\alpha - 2^{nd} fit)$ provided by the fit of the background subtracted spectrum. Since we cannot determine the fractions of the narrow H$\alpha$ component due to the star and due to the nebular emission, it is not possible to calculate the emission-line ratios [N II] $\lambda 6584$/H$\alpha$ and ([S II] $\lambda 6716 + \lambda 6731$)/H$\alpha$ for region 4, without a considerable uncertainty. Therefore, we decided to not include the point representing region 4 in the diagnostic diagrams in Fig.~\ref{fig6}.   

In order to analyse the kinematics of the ionized gas in the nuclear region of NGC 1313, we determined the radial velocity and the velocity dispersion of the gas ($V_{gas}$ and $\sigma_{gas}$, respectively) of each spaxel of the data cube by fitting a Gaussian function to the H$\alpha$ emission line of the corresponding spectrum. The amplitude/noise (A/N) ratio of the H$\alpha$ emission line is $\sim 20$ in the most peripheral areas of the FOV and is larger than 600 near the stellar nucleus. Such large values of the A/N ratio allowed a precise analysis of the kinematic parameters. However, the analysis of the spectra in the area corresponding to region 4 is more complicated. Since we are performing this kinematic analysis using simple Gaussian fits, we could not reproduce properly the profile of the H$\alpha$ line in the spectra of region 4 (where fits involving two Gaussians are required to reproduce the H$\alpha$ profile, as explained before). As a consequence, the values of $\sigma_{gas}$ obtained for that area were not reliable. On the other hand, we verified that the fit of only one Gaussian to the H$\alpha$ line in the spectra of region 4 provided, at least, precise values for $V_{gas}$. Fig.~\ref{fig8} shows the obtained $V_{gas}$ and $\sigma_{gas}$ maps. The non-reliable values of $\sigma_{gas}$ in region 4 were masked in the corresponding map. 

The $V_{gas}$ map reveals a pattern consistent with rotation, although a number of irregularities can be easily seen. The line of nodes has a position angle (PA) of $10\degr \pm 10\degr$. We assumed that the centre of the $V_{gas}$ map (highlighted in Fig.~\ref{fig8}) is given by the point, along the line of nodes, where the velocity is equal to the average between the maximum and minimum values ($V_{average}$). The $V_{gas}$ map shows the values obtained after the subtraction of $V_{average}$ or, in other words, the values of $V_{gas}$ relative to the centre of the map. The moduli of the radial velocities are lower than 25 km s$^{-1}$ along the entire FOV. The systemic velocity of NGC 1313 was taken as being equal to the sum of $V_{average}$ and the velocity corresponding to the redshift (obtained from NED) used to pass the original data cube to the rest frame, which resulted in $V_{sys} = 475 \pm 6$ km s$^{-1}$. The centre of the $V_{gas}$ map in Fig.~\ref{fig8} is not exactly coincident with the centre of region 1, although these two positions are compatible at $1\sigma$ level. The projected distance between the stellar nucleus and the centre of the $V_{gas}$ map is $0.31 \pm 0.06$ arcsec (which corresponds to $6.8 \pm 1.3$ pc). Regions 2 and 3 are both in blueshift, with similar velocities, while region 4 is in redshift. 

\begin{figure*}
\begin{center}
  \includegraphics[scale=0.50]{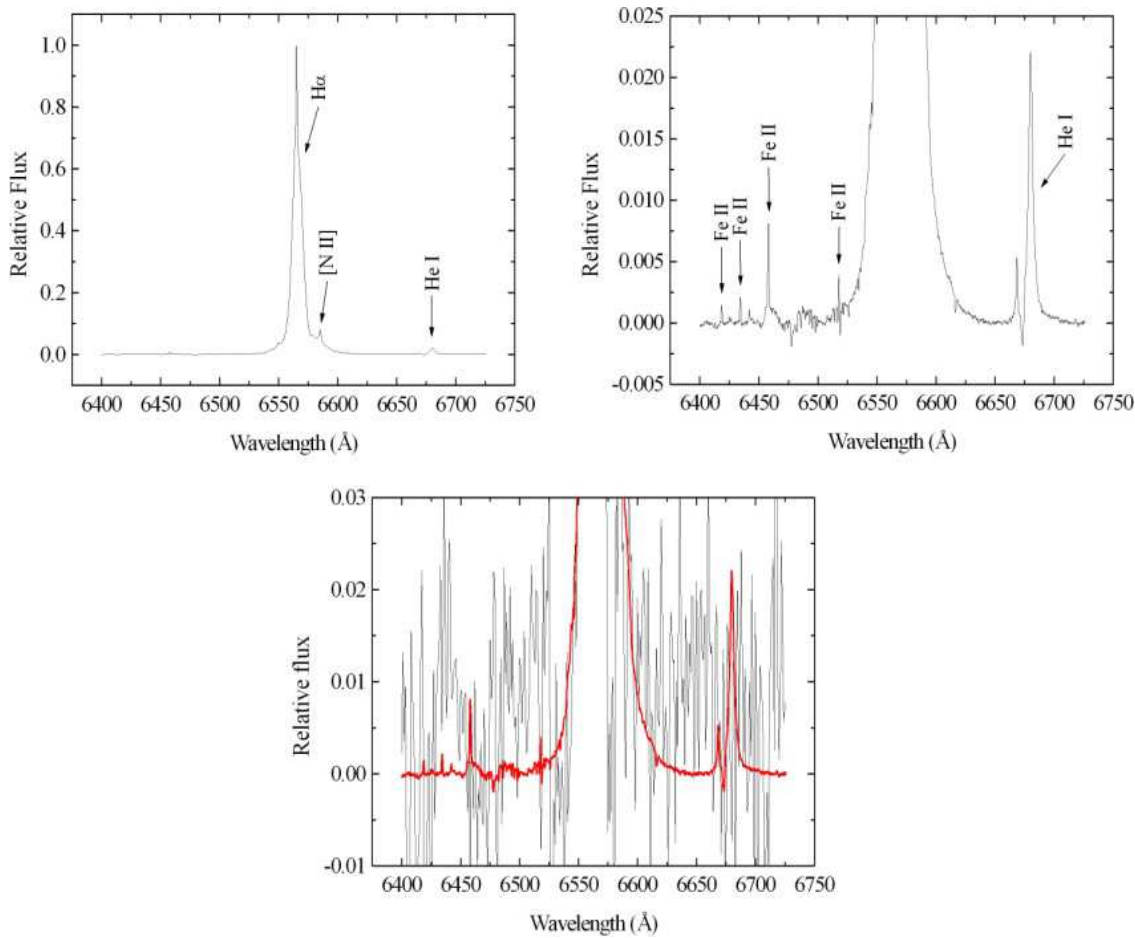}
  \caption{Top left: optical spectrum of the LBV $\eta$ Carinae. Top right: magnification of the spectrum shown at top left. Bottom: superposition of the spectrum of $\eta$ Carinae (shown in red) with the spectrum of region 4 (shown in black). All the spectra were scaled in order to make the strength of their H$\alpha$ lines equal to 1.\label{fig10}}
\end{center}
\end{figure*}

The $\sigma_{gas}$ map is also very irregular, but we can see that the dispersion values (corrected for the spectral resolution) appear to be lower at the area where the rotation pattern is observed. In particular, the lowest values of $\sigma_{gas}$ were detected very close to the centres of region 1 and of the $V_{gas}$ map, originating what we can call a $\sigma_{gas}$-drop. The dispersion values are in the range of $10 - 43$ km s$^{-1}$ along the FOV. Fig.~\ref{fig8} also shows curves extracted, along the line of nodes, from the $V_{gas}$ and $\sigma_{gas}$ maps. The uncertainties of the values were obtained by performing a Monte Carlo simulation and also taking into account the spectral resolution of the observation.

\section{Discussion and comparison with previous studies}

\subsection{The WR(s) in region 1}

WRs are evolved, central He-burning, massive stars that produce intense stellar winds, which result in considerable mass-loss. Such stellar winds affect significantly the surrounding medium by releasing mechanical energy and nuclear processed material. The study of WRs is of great importance, as these objects are believed to be the progenitors of type Ib/c supernovae and long gamma-ray bursts (for more information, see Crowther 2007 and references therein). Spectroscopically, WRs are characterized by broad emission lines of He, N (WNs), C (WCs) or O (WOs). Because of its easily identifiable properties, WRs have been detected in the Milky Way (e.g. Crowther et al. 2006; Fern\'andez-Mart\'in et al. 2012; Shara et al. 2012; Smith et al. 2012; Kanarek et al. 2015) and in nearby galaxies (e.g. Massey \& Johnson 1998; Castellanos, D\'iaz \& Terlevich 2002; Massey 2003; Hadfield \& Crowther 2006; Eldridge \& Rela\~no 2011; Neugent \& Massey 2011; Bibby \& Crowther 2012; Howarth \& Walborn 2012; Neugent, Massey \& Georgy 2012; Neugent, Massey \& Morrell 2012; Kehrig et al. 2013; Shara et al. 2013; Gvaramadze et al. 2014a; Massey, Neugent \& Morrell 2015; Sokal et al. 2015). 

The presence of WRs close to the nucleus of NGC 1313 was detected by \citet{had07}, in an H II region identified by the authors as 37. We believe that this H II region corresponds to region 1 in our data cube. The red bump around $5808 \AA$ in the spectrum of region 1 is characteristic of WCs. A sub-classification of WCs is based on the C IV $\lambda 5808$/C III $\lambda 5696$ ratio \citep{smi90}. Since the C III $\lambda 5696$ lines are not visible in the spectrum of region 1, we believe that a classification of WC4-5 is appropriate. \citet{had07} obtained the same classification. Taking $L_{WC4}(5808) = 3.0 \times 10^{36}$ erg s$^{-1}$ \citep{sch98} as a typical luminosity for the red bump (containing the C IV $\lambda 5808$ lines) of a WC4, we conclude that the luminosity of the red bump in the spectrum of region 1 ($L_{red~bump} = (5.17 \pm 0.26) \times 10^{36}$ erg s$^{-1}$) suggests the existence of only one or two WCs in this area, which is consistent with the result obtained by \citet{had07}.

\subsection{The emission-line star in region 4}

As mentioned before, the spectral features of region 4 suggest the presence of a massive emission-line star there. We retrieved an \textit{HST} image of NGC 1313, obtained with the Wide-field Camera 3 (WFC3) in the F336W filter (\textit{U} band). Fig.~\ref{fig9} shows a comparison between the image of the data cube of NGC 1313, before the starlight subtraction, collapsed along the spectral axis, the H$\alpha$ image of the data cube, after the starlight subtraction, and also the \textit{HST} image, with the same FOV of the data cube. The LBV/B[e] supergiant/B hypergiant candidate appears very clearly in the \textit{HST} image, which also shows other hot stars along the FOV. The two brightest structures in the \textit{HST} image (one coincident with the nucleus and the other $0.62 \pm 0.05$ arcsec south from the nucleus) appear as a single elongated area in the image of the collapsed GMOS data cube and are probably associated with regions 1 and 3 in the H$\alpha$ image.

A natural question at this point is: what is the nature of the emission-line star in region 4 ? There are different possibilities here. A B[e] star, for example, is a spectral type B star with forbidden emission lines in its spectrum. It is established that a star with (a) strong Balmer lines, (b) low excitation permitted emission lines (e.g. Fe II), (c) forbidden optical emission lines of [Fe II] and [O I] and (d) strong near- or mid-infrared excess due to hot circumstellar dust shows the B[e] phenomenon \citep{zic98}. Other properties like luminosity, association with star forming regions or the presence of inverse P Cygni profiles can be used to classify B[e] stars as B[e] supergiants, pre-main sequence B[e]s, etc. \citep{lam98}. 

An LBV, on the other hand, is an evolved massive star that shows considerable spectral and photometric variability on time-scales from months to years. At minimum visual brightness (the ``quiescent state''), the spectrum of an LBV is similar to that of a hot supergiant, while, at maximum visual brightness (the ``eruptive state''), an LBV has an A-F supergiant spectrum. The emission-line spectra of LBVs show prominent H, He I, Fe II and [Fe II] lines, sometimes with P-Cygni profiles (for more details, see Humphreys \& Davidson 1994). One explanation for the LBVs is that these objects are in a transition phase between the main sequence and the WR phase \citep{hum94,gro14}. Another possibility is that these stars are in the final phase before supernova explosion (Groh, Meynet \& Ekstr\"{o}m 2013). The spectra of B[e] supergiants and LBVs in quiescent state are actually very similar to each other, although it is still not clear if there is a relation between these two types of stars \citep{kra14}. So far, a set of B[e] and LBV candidates have been found in the Milky Way (e.g. Humphreys \& Davidson 1994; Gvaramadze et al. 2010; Clark, Ritchie \& Negueruela 2013; Gvaramadze et al. 2014b; Liermann et al. 2014;) and in nearby galaxies (e.g. Humphreys \& Davidson 1994; Fabrika et al. 2005; Massey et al. 2007; Valeev, Sholukhova \& Fabrika 2009, 2010; Graus, Lamb \& Oey 2012; Humphreys et al. 2014; Kalari et al. 2014; Levato, Miroshnichenko \& Saffe 2014; Sholukhova et al. 2015). 

Another group of emission-line stars is composed by the B hypergiants, which also show Balmer emission lines. There are similarities between the spectra of B hypergiants and LBVs; however, the spectra of the formers usually have weaker H I and He I lines and also do not show low excitation metallic lines, like Fe II and Fe III, which are characteristic of LBVs (for more details about B hypergiants, see Clark et al. 2012). 

The spectrum of region 4 has, at least, one of the characteristics of the B[e] phenomenon: a strong H$\alpha$. The EW of the broad component of H$\alpha$ (before and after the background subtraction) is consistent with the values expected for an LBV (e.g. Clark \& Negueruela 2004). We can only conclude that the object in region 4 is an LBV/B[e] supergiant/B hypergiant candidate.

In order to evaluate whether we have enough S/N to detect possible Fe II or He I lines in the spectrum of region 4, first of all, we retrieved the optical spectrum of a well known LBV, $\eta$ Carinae, which was taken by A. Damineli, on 2016 June 15, with the Coud\'e spectrograph of the 1.6 m telescope at Observat\'orio do Pico dos dias of the Laborat\'orio Nacional de Astrof\'isica, in Brazil.  That spectrum has a strong broad H$\alpha$ line, but Fe II and He I lines are also visible. We scaled the spectra of region 4 (after the background subtraction) and of $\eta$ Carinae in order to make the strength of the H$\alpha$ line in both of them to be equal to 1. Fig.~\ref{fig10} shows the scaled spectrum of $\eta$ Carinae and also the superposition with the scaled spectrum of region 4. It is easy to see that the intensities of all the Fe II and He I lines are lower than the noise in the spectrum of region 4. Therefore, based on this comparison, we conclude that, if the emission-line star in region 4 is an LBV like $\eta$ Carinae, our S/N is not enough to detect the Fe II  or the He I emission lines, which are significantly weaker than H$\alpha$.  

Without the detection of Fe II or He I lines in the spectrum, it is really difficult to obtain a classification of the emission-line star in region 4. Nevertheless, there is another property that should be evaluated: the spectral and photometric variability. Unlike LBVs, B[e] stars and B hypergiants do not show long-term spectral and photometric variability. Although we do not have observations to properly monitor any spectral or photometric variability of the object in region 4, we performed a simple test, using another \textit{HST} image, obtained with ACS in the F555W filter (\textit{V} band). First, we reproduced an image in the ACS-F555W filter using the GMOS data cube. To do that, we obtained an image of the integrated flux of the data cube, taking into account the response curve of the ACS-F555W filter. After that, we convolved the \textit{HST} image with an estimate of the PSF of our GMOS-IFU observation. Then, we defined two circular regions, with radii of 0.4 arcsec, in the \textit{HST} image and in the integrated GMOS-IFU image: one of them centred on region 4 and the other centred on the stellar nucleus of NGC 1313. Finally, for both images, we calculated the difference of magnitude between these two circular regions: $\Delta m = m(4) - m(nucleus)$. Since the \textit{HST} image was taken on 2004 July 17, a possible variability of the emission-line star in region 4 could be revealed by this test. The values obtained for the integrated GMOS-IFU image and for the \textit{HST} image, respectively, were $\Delta m(GMOS-IFU) \sim 0.31$ mag and $\Delta m(HST) \sim 0.35$ mag. These two values are very similar to each other and do not show any variability of the source in region 4. The absence of such variability would be consistent with the hypothesis of a B[e] supergiant or a B hypergiant in this area. However, the fact that this simple test did not detect a variability, of course, does not prove that the source in region 4 is not variable and a longer monitoring is required.

\subsection{The kinematically cold nucleus}

The PA we obtained for the line of nodes of the $V_{gas}$ map of the H$\alpha$ emission line ($10\degr \pm 10\degr$) is compatible, at 1$\sigma$ level, with the value obtained by Marcelin \& Athanassoula (1982, $-10\degr \pm 10\degr$) from an analysis of interferograms in H$\alpha$ of NGC 1313. Our value is also compatible, at 2$\sigma$ level, with the one obtained by Peters et al. (1994, $-6\degr \pm 3\degr$) from an analysis of the H I velocity field of this galaxy. On the other hand, there is a higher discrepancy between the PA obtained by Carranza \& Ag\"{u}ero (1974, $-20\degr$), also from analysis of interferograms in H$\alpha$, and by us, although these two values are still compatible, at $3\sigma$ level. The differences between the PA of the line of nodes obtained by us and by previous studies were probably caused by the fact that we analysed here a much smaller FOV, which usually provides a more imprecise determination of this PA. In addition, most of these previous studies detected significant corrugations in the isovelocity contours of this galaxy, which may result in variations of the exact value of the PA discussed here along the galaxy. The irregularities we observed in the $V_{gas}$ map shown in Fig.~\ref{fig8} were actually expected, as there are hot massive stars along the FOV (including WRs), which can produce winds (and undergo significant mass-loss) that could result in the observed irregularities. 

The centre of the $V_{gas}$ map is displaced by $0.31 \pm 0.06$ arcsec ($6.8 \pm 1.3$ pc), to the north-west, from the stellar nucleus. \citet{pet94} verified that the centre of their kinematic map of H I was displaced by 20 arcsec $\pm$ 15 arcsec to the east and 0 arcsec $\pm$ 15 arcsec to the north from the stellar nucleus; however, based on the errors of their measurements, the authors concluded that the kinematic centre was essentially coincident with the stellar nucleus. In our case, the centre of the $V_{gas}$ map is not compatible with the stellar nucleus, even at $3\sigma$ level, but, considering the small projected distance between these two positions, the irregularities in our $V_{gas}$ map and also the much smaller FOV we are analysing (which can lead to an imprecise determination of the kinematic centre), we can say that our results are consistent with the ones obtained by \citet{pet94}. 

The $\sigma_{gas}$ map revealed some interesting features. The lower values of $\sigma_{gas}$ along the area where the rotation pattern was detected suggests the existence of a cold gaseous disc in rotation around the nucleus. The presence of an apparent $\sigma_{gas}$-drop deserves special attention. Drops in the values of the stellar velocity dispersion ($\sigma_{stellar}$-drop) in the nuclear regions of galaxies have been increasingly observed in the last years (e.g. Emsellem et al. 2001; de Zeeuw et al. 2002; M\'arquez et al. 2003; Shapiro, Gerssen \& van der Marel 2003; Falc\'on-Barroso et al. 2006; Ganda et al. 2006). The most accepted scenario to explain this phenomenon in disc galaxies is that it is associated with a stellar population in the nuclear region of the galaxy formed from a circumnuclear cold gaseous disc. The recently formed stars keep the low velocity dispersions from the gas clouds from which they formed, which results in a kinematically cold young stellar population in the nuclear region of the galaxy. The flux in the observed spectrum is dominated by the emission and, therefore, by the low velocity dispersion values from the young stellar populations, originating the so-called $\sigma$-drop \citep{woz03,all05,all06}. \citet{fal06} analysed SAURON data cubes of the central regions of 24 spiral galaxies and verified that the objects with higher star formation rates have lower $\sigma_{gas}$ values, indicating that the cold gas is forming stars. We believe that all these scenarios are applicable to the nuclear region of NGC 1313. The position of the observed $\sigma_{gas}$-drop is almost coincident with the position of region 1, which is an H II region with a probable high number of young massive stars. Therefore, a possible explanation is that the stars in region 1 were actually formed by this cold gas. Although the weak absorption spectra in the data cube of NGC 1313 did not allow reliable stellar kinematic measurements, we believe that, if this hypothesis is correct, low stellar velocity dispersion values are also expected for region 1.

\section{Summary and conclusions}

We performed the first detailed analysis of the line-emitting areas in the nuclear region of NGC 1313, using an optical GMOS-IFU data cube. After a carefull starlight subtraction, four main emitting areas were identified in an image of the H$\alpha$ emission-line. The spectrum of region 1 shows broad emission features around $4650 \AA$ (representing a blend of N III, C III, C IV and He II lines) and $5808 \AA$ (representing a blend of C IV lines), which are typical of WRs. Our analysis revealed the presence of one or two WC4-5 stars in this region. The same result was obtained by \citet{had07}. 

The spectrum of region 4 has, at least, one of the characteristics of the B[e] phenomenon: a strong H$\alpha$ with a broad component, which indicates the presence of an emission-line star there. A comparison with a \textit{V} band \textit{HST} image of this galaxy showed no apparent variability of region 4. However, a proper monitoring is required, in order to evaluate the possibility of photometric and spectral variability. We also verified that, if the emission-line star is similar to the well known LBV $\eta$ Carinae, then our S/N is not enough to detect possible He I or Fe II lines in the spectrum. Therefore, based on that, we can only conclude that there is an LBV/B[e] supergiant/B hypergiant candidate in region 4, but further analysis is required, in order to properly classify this star. Diagnostic diagram analysis revealed that the emission-line ratios of regions 1, 2 and 3 are consistent with those of H II regions. 

The $V_{gas}$ map we obtained, based on the H$\alpha$ emission line, revealed a rotation pattern. Although we are analysing a considerably small FOV, the PA of the line of nodes we obtained for that map is compatible (at $1\sigma$, $2\sigma$ or $3\sigma$ level) with the values obtained by previous studies. We verified that the centre of the $V_{gas}$ map is located at a projected distance of $0.31 \pm 0.06$ arcsec ($6.8 \pm 1.3$ pc), to the north-west, from the stellar nucleus. The values of $\sigma_{gas}$ are significantly lower along the area where the rotation pattern was detected, suggesting the presence of a cold rotating gaseous disc. In addition, we detected a $\sigma_{gas}$-drop almost coincident with the position of region 1, which suggests that the young massive stars there were probably formed by this cold gas, possibly keeping low values of velocity dispersion. Therefore, we predict that this galaxy may also show a $\sigma_{stellar}$-drop in its nuclear region.

\section*{Acknowledgements}

This work is based on observations obtained at the Gemini Observatory (processed using the Gemini IRAF package), which is operated by the Association of Universities for Research in Astronomy, Inc., under a cooperative agreement with the NSF on behalf of the Gemini partnership: the National Science Foundation (United States), the National Research Council (Canada), CONICYT (Chile), the Australian Research Council (Australia), Minist\'{e}rio da Ci\^{e}ncia, Tecnologia e Inova\c{c}\~{a}o (Brazil) and Ministerio de Ciencia, Tecnolog\'{i}a e Innovaci\'{o}n Productiva (Argentina). This work was supported by CAPES and FAPESP (under grants 2012/02268-8 and 2011/51680-6). We thank Dr Jos\'e Groh and an anonymous referee for valuable comments on the paper. We also thank Dr Augusto Damineli for providing us the optical spectrum of $\eta$ Carinae and also for important comments on this paper.

\end{document}